\documentclass[12pt,leqno]{article}
\usepackage{amsmath,amsthm,amssymb}

\def\init{\setcounter{equation}{0}}
\newtheorem{theorem}{Theorem}[section]
\newcommand{\R}{\mathbb{R}}

\newcommand{\Z}{\mathbb{Z}}

\newcommand{\e}{{\varepsilon}}

\newcommand{\rw}{\rightarrow}

\usepackage{tikz}
\usetikzlibrary{decorations.pathreplacing}

\usepackage{verbatim}

\title{Hawking radiation  from acoustic black holes  in two  space  dimensions }

\author{G.Eskin, \ \ \  Department of Mathematics, UCLA,\\ Los Angeles,
CA 90095-1555, USA. \ E-mail: eskin@math.ucla.edu
}

\begin{document}

\maketitle

\begin{abstract}
We study  the Hawking radiation  for the acoustic black hole.
In the beginning we follow the outline of T.Jacobson  but then we use a different $2+1$ vacuum state  similar to the vacuum state 
 constructed   by W.Unruh.
We  also  use  a special  form  of the wave packets.  The focus of the paper is to treat  the 2 dimensional case,  in particular,  the case  when 
 the radial  and angular  velocity  are variable.  
\end{abstract}

\section{Introduction}
\init

Consider  the acoustic  wave equation  of the form
\begin{equation}																		\label{eq:1.1}
\Box_g u(x_0,x)=\frac{1}{\sqrt{g(x)}}\sum_{j,k=0}^2\frac{\partial}{\partial x_j}\Big(\sqrt g \, g^{jk}(x)\frac{\partial u(x_0,x)}{\partial x_k}\Big)=0,
\end{equation}
where  $x_0\in \R$  is  the time variable,  $x=(x_1,x_2)\in \R^2,\ g(x)=(\det[g^{jk}]_{j,k=0}^2)^{-1},$
\begin{align}																		\label{eq:1.2}
&g^{00}=1,\ \ g^{0j}=g^{j0}=v^j,\ \ \  1\leq j\leq 2,
 \\
\nonumber
&g^{jk}=(-\delta_{ij}+v^jv^k),\ \ \ 1\leq j,k\leq 2,
\end{align}
$v=(v_1,v_2)=\frac{A}{\rho}\hat x +\frac{B}{\rho}\hat \theta$  is  the velocity  flow  in a vortex,  $\rho=|x|,\hat  x=\frac{x}{|x|},
\linebreak
\hat\theta 
=\big(-\frac{x_2}{|x|},\frac{x_1}{|x|}\big)$.  We assume,
for the simplicity,  that the density and the sound speed  are equal to 1.  We assume also that
\begin{equation}																	\label{eq:1.3}
A<0,  \ \  B\neq 0.
\end{equation}

The equation  (\ref{eq:1.1}) describes  the wave propagation in the moving fluid
with velocity $(v_1,v_2)=\frac{A}{\rho}\hat x +\frac{B}{\rho}\hat \theta$  (cf. [27]).
The metric  corresponding to  (\ref{eq:1.1})  has the form
\begin{equation}                                                                                                   						\label{eq:1.4}
ds^2=dx_0^2-(dx-vdx_0)^2,
\end{equation}
where $x=(x_1,x_2),v=(v_1,v_2)$.

It is a Lorentzian metric  in $\R^2\times\R$   called the acoustic metric.   It is  one of many  examples  of analogue gravity metrics  
(see,  for example,  the survey [2])   where many others physical  examples  of analogue  gravity  are given).

In the general relativity  the metric  is a solution  of the Einstein  equation  and this  sets the analogue   metrics  apart.  From other side analogue
metrics exhibit  many  properties of metrics  in general   relativity,  in particular,  they  both may have black holes.  This  fact  spurred  the interest
of physicists in the theoretical and experimental studies  of analogue  gravity (cf.  [16],  [22], [26],  [27]).

The black hole,  by the definition,  is  a domain  in a spacetime where the signals or particles can not escape  
from.
It was a remarkable  discovery of S.Hawking  that  when  quantum  effects  are added,  the black  hole emits particles  ([13]).
This phenomenon is called the Hawking radiation  and there is a large literature  devoted to this subject  (cf. [1],  [4],  [5],  [12],  [16],  [17],  [26]).  
Hawking  radiation  holds for  analogue  black  holes too and  this ignites  an additional   interest  in  Analogue Gravity.  Note  that  it is possible  to 
demonstrate  
the Hawking  radiation  for analogue black  holes  while the experimental  verification of Hawking  radiation  in general  relativity  is not  realistic.  The pioneering work  in this direction  was done  by W.Unruh  in [22].   For more  recent  experimental  results  see [3], [10], [20],  [21],  [29].

Hawking  radiation from  acoustic black holes  was considered  in [25],  [30].

In all preceding works on  Hawking  radiation  the case of $1+1$ dimensions or  spherically symmetric  case  in  $3+1$  dimensions
were studied.

The focus  of this paper  is to investigate the Hawking radiation in  $2+1$  dimensions,  primary the  rotating 
acoustic black holes  with  variable
radial and angular velocities.

Note that  classical  acoustic  black hole  with the variables velocity were treated in  [6],  [7],  [18],  [19],  [28].

As an introduction  to the quantum  field theory  on curved  spacetimes we  use the lecture notes of T.Jacobson  [15] 
 (see  also  [14]  and references there).

The main departure from  [15]  in this paper is a new  definition of  the vacuum  state  that  follows        the definition  of the  Unruh  vacuum  
(cf.  [23],  [24]).

Another novelty  is
 a special construction  of wave packets  that is used  in the computation  of the  Hawking  radiation.  For comparison, 
K.Fredenhagen  and  R.Haag  [11]   in one  of   few  rigorous   derivation    of  Hawking  radiation  used  a limiting procedure  
when
$x_0\rw -\infty$  
to obtain  the expression  for the Hawking radiation.  Using  special  wave  packets  we avoid  taking the time limit. 
Instead,  we  use  the limit  $a\rw\infty$  of some parameter  $a$   that  characterizes the closeness  of the wave  packet to the black hole.

The plan of the paper is  the following:

In \S 2 we describe    needed facts from the quantum  field  theory  on curved  spacetimes  and  define  the new   vacuum state.

In \S3   we study  a more  simple case  of the acoustic  black hole  when  $A< 0$  and  $B\neq  0$  are constant.  We  introduce a wave packets  of 
a special form  and compute the  Hawking radiation produced  by such wave packets.

Finally,  in  \S 4 we consider  the acoustic  black holes  with the  variable  velocity.  In \S 4.1 we study  the Hawking  radiation  in the case
when  $A<0$  is constant  and  $|B(\varphi)|>0$   is periodic in  $\varphi$   and  in  \S 4.2  we study  the case  of general acoustic black hole 
 where  $A(\rho,\varphi)<0, |B(\rho,\varphi)|>0$.

Some  $A(\rho,\varphi),  B(\rho,\varphi)$    have direct physical  meaning (cf.  [27]).   In order  for  
$\frac{A}{\rho}\hat x +\frac{B}{\rho}\hat \theta$ 
 to represent  a fluid  flow there should be  a harmonic  function  $\Psi(\rho,\varphi)$  such that
$A=\rho\frac{\partial\Psi}{\partial r}$ and $B=\frac{\partial\Psi}{\partial \varphi}$  (see [27]).  
 If we take  $\Psi= A_0\log \rho + B_0\varphi
+ C_1 \rho\cos\varphi + C_2 \rho\sin\varphi$,
then  $A(\rho,\varphi)=A_0+C_1 \rho\cos\varphi +C_2 \rho\sin\varphi,\ B(\rho,\varphi)= B_0-C_1 \rho\sin\varphi + C_2\rho \cos\varphi$
(cf.  [7],  \S 4). 

Note that one can take  any  harmonic  polynomials instead  of $C_1\rho\cos\varphi +C_2 \rho\sin\varphi$.

\section{Elements  of the second quantization}
\init

For the elements of the quantum  field  theory on the curved space-time used in this paper see the lectures notes  of  T.Jacobson  [15]  (see also  [14]
  and the further 
references  there).

Denote  by  $f_k^+(x_0,x) $  the  solution  of   $\Box_g u =0$   in   $\R^2\times\R$  with 
the initial  conditions having  the following   form  in polar coordinates  $(\rho,\varphi)$:
\begin{equation}															\label{eq:2.1}
f_k^+(x_0,x)\big|_{x_0=0}=\gamma_k e^{ik\cdot x},
\end{equation}
where $k=(\eta_\rho,\eta_\varphi), x=(\rho,\varphi), k\cdot x=\eta_\rho\rho+\eta_\varphi\varphi$,
\begin{equation}														\label{eq:2.2}
\frac{\partial f_k^+}{\partial x_0}\big|_{x_0=0}=i\lambda_0^-(k)\gamma_ke^{ik\cdot x},
\end{equation}
where
\begin{equation}														\label{eq:2.3}
\lambda_0^-(k)=-\frac{A}{\rho}\eta_\rho-\frac{B\eta_\varphi}{\rho^2}-\sqrt{\eta_\rho^2+\frac{\eta_\varphi^2}{d^2}},\ \ \ d \ \ \mbox{is arbitrary}.
\end{equation}
The normalization  factor $\gamma_k$  will  be chosen later. 
Also  denote  by  $f_k^-(x_0,x)$  the solution of  $\Box_g u=0$   with  the initial  condition 
\begin{equation}														\label{eq:2.4}
f_k^-(0,x)=\gamma_k e^{ik\cdot x},\ \ \  \frac{\partial f_k^-}{\partial x_0}\Big|_{x_0=0}=i\lambda_0^+(k)\gamma_k e^{ik\cdot x},
\end{equation}
where $\lambda_0^+(k)=-\frac{A}{\rho}\eta_\rho-\frac{B}{\rho^2}\eta_\varphi+\sqrt{\eta_\rho^2+\frac{\eta_\varphi^2}{d^2}}$.
Note that 
\begin{equation}														\label{eq:2.5}
\overline{f_k^+}(x_0,x)=f_{-k}^-(x_0,x),
\end{equation}
since $-\lambda^+(\rho,\eta_\rho,\eta_\varphi)=\lambda^-(\rho,-\eta_\rho,-\eta_\varphi)$   and  $\gamma_k=\gamma_{-k}$ 
is positive.
 
 Let
 \begin{equation}														\label{eq:2.6}
 <f,h>=i\int\limits_{x_0=t}g^{\frac{1}{2}}\sum_{j=0}^2 g^{0j}\Big(\overline f\frac{\partial h}{\partial x_j}-
 \frac{\partial \overline f}{\partial x_j}h\Big)dx_1dx_2.
 \end{equation}
The bracket (\ref{eq:2.6})  is called  the Klein-Gordon  (KG)  inner product.  
Since  the acoustic  metric  (\ref{eq:1.4})   is  stationary,  the inner product (\ref{eq:2.6})  is independent  of $t$
when  $f, h$  are solutions of $\Box_g u=0$  (cf.  [15]).
 We have, 
 in $(\rho,\varphi)$ coordinates:
\begin{align}	
\nonumber
&<f_k^+,f_{k'}^+>
=
i\int\limits_{x_0=0}
\Big[\Big(\overline{f_k^+}
\frac{\partial f_{k'}^+}{\partial x_0}-\frac{\partial \overline{f_k^+}}{\partial x_0}f_{k'}^+\Big)
+\frac{A}{\rho}\Big(\overline{f_k^+}\frac{\partial f_{k'}^+}{\partial \rho}-\frac{\partial \overline{f_k^+}}{\partial \rho}f_{k'}^+\Big)
\\
\nonumber
&+\frac{B}{\rho^2}\Big(\overline{f_k^+}\frac{\partial f_{k'}^+}{\partial \varphi}-\frac{\partial \overline{f_k^+}}{\partial \varphi}f_{k'}^+\Big)\Big]
\rho\, d\rho\, d\varphi     
\\
\nonumber
&=
i\int\limits_{x_0=0}\overline{f_k^+} f_{k'}^+\Big(i\lambda_0^-(k')+i\frac{A}{\rho}\eta_\rho'+i\frac{B}{\rho^2}\eta_\varphi'
+i\lambda_0^-(k)+i\frac{A}{\rho}\eta_{\rho}
+i\frac{B}{\rho^2}\eta_\varphi\Big)\rho\,d\rho\,d\varphi,
\end{align}
where
$k=(\eta_\rho,\eta_\varphi),k'=(\eta_\rho',\eta_\varphi')$.               
Note that $\gamma_k$  is real-valued.  Then  derivatives  of $\gamma_k$  do not contribute  to KG norm.
Also note that  $\lambda_0^-(k)+\frac{A}{\rho}\eta_\rho+\frac{B}{\rho^2}\eta_\varphi=-\sqrt{\eta_\rho^2+\frac{1}{d^2}\eta_\varphi^2}$.
Therefore
\begin{equation}														\label{eq:2.7}
<f_k^+,f_{k'}^+>=\int\limits_{x_0=0}\Bigg(\sqrt{\eta_\rho^2+\frac{\eta_\varphi^2}{d^2}}+
\sqrt{(\eta_\rho')^2+\frac{(\eta_\varphi')^2}{d^2}}\Bigg)\gamma_k \gamma_{k'}\, e^{-i(k-k')\cdot x}\rho\,d\rho\,d\varphi.
\end{equation}
We choose a
normalizing factor $\gamma_k=\frac{1}{\sqrt{\rho}\big(\eta_\rho^2+\frac{\eta_\varphi^2}{d^2}\big)^{\frac{1}{4}}{\sqrt{2(2\pi)^2}}
}$.
Then
$<f_k^+,f_{k'}^+>=\frac{1}{(2\pi)^2}\int\limits_{\R^2} e^{-i(k-k')\cdot x}\,d\rho\,d\varphi=\delta(k-k')$.
Analogously,
\begin{equation}														\label{eq:2.8}
<f_k^-,f_{k'}^->=-\delta(k-k'),\ \ \ 
<f_k^+,f_{k'}^-> =0.
\end{equation}
Note that  $\varphi\in [0,2\pi]$  and  $\eta_\varphi=m\in\Z$.  Therefore  $\int_{-\infty}^\infty(\ \ )d\eta_\varphi$  is,    in fact,  a sum  
$\sum_{m=-\infty}^\infty$
and  $\delta(\eta_\varphi-\eta_{\varphi'})=\delta_{mm'},\ \delta_{mm'}=1$  when  $m=m'$  and  $\delta_{mm'}=0$  when    $m\neq  m'$.

We shall  continue   to use  $\eta_\varphi\in\R$  as  in  (\ref{eq:2.1})-(\ref{eq:2.4})  simply  for the  shortness of  notations.

Let $\Phi$  be the  field operator, i.e.
\begin{equation}														\label{eq:2.9}
\Phi=\int\limits_{\R^2}(\alpha_k^+f_k^+(x_0,x)-\alpha_{-k}^-f_{-k}^-(x_0,x))dk,
  \end{equation}
  where                                                                                                                                                                                                                                                                                                                                                                                                                                                                                                                                                                                                                                                                                                                                                                                                                                                                                                                                                                                                                                                                                                                                                                                                                                                                                                                                                                                                                                                                                                                                                                                                                                                                                                                                                                                                                                                                                                                                                                                                                                                                                                                                                                                                                                                                                                                                                                                                                                                                                                                                                                                                                                                                                                                                                                                                                        \begin{equation}														\label{2.10}
  \alpha_k^+=<f_k^+,\Phi>,\ \ \alpha_{-k}^-=(\alpha_k^+)^*=<f_{-k}^-,\Phi>
  \end{equation}
  are  the  annihilations and creations operators,  respectively.
  
  It follows  from  (\ref{eq:2.7}),  (\ref{eq:2.8})   (cf.  [15])   that  operators  $\alpha_k^+$  and  $\alpha_{-k}^-$  satisfy the following  
  commutation  relations
  \begin{align}															\label{eq:2.11}
  &[\alpha_k^+,\alpha_{-k'}^-]=\delta(k-k')I,\ \ \ [\alpha_k^+,\alpha_{k'}^+]=0,
  \\
  \nonumber
   &   [\alpha_{-k}^-,\alpha_{-k'}^-]=0,
   \ \ \
   \alpha_{-k}^-=(\alpha_{k}^+)^*.                                                                                                                                                                                                                                                                                                                                                                                                                                                                                                                                                                                                                                                                                                                                                                                                                                                                                                                                                                                                                                                                                                                                                                                                                                                                                                                                                                                                                                                                                                                                                                                                                                                                                                                                                                                                                                                                                                                                                                                                                                                                                                                                                                                                                                                                                                                                                                                                                                                                                                                                                                                                                                                                                                                                                                                                                                                                                                                                                                                                                                                                                                                                               
  \end{align}
  Let  $C(x_0,\rho,\varphi)$  be  a solution  of  (\ref{eq:1.1})   with  initial  conditions at  $x_0=0$  having  a   support  in
  the exterior  of  the black  hole  $\{\rho>|A|\}$.   We shall call $C$  a wave  packet.
  
  Expanding  $C$  in a basis $f_k^+(x_0,x),f_{-k}^-(x_0,x)$  we have
  \begin{equation}														\label{eq:2.12}
  C=\int\limits_{\R^2}\big(C^+(k)f_k^+(x_0,x)-C^-(k)f_{-k}^-(x_0,x)\big)dk,
  \end{equation}
  where 
  $
  dk=d\eta_\rho d\eta_\varphi$. 
  As it was mentioned  above  the integration  in $\eta_\varphi$  is indeed a summation  $\sum_{m=-\infty}^\infty$.
   In  (\ref{eq:2.12})
  \begin{equation}														\label{eq:2.13}
  C^+(k)=<f_k^+,C>,\ \ \ C^-(k)=<f_{-k}^-,C>.
  \end{equation}
  Note  that  integrals  in (\ref{eq:2.9})  are  understood in a distribution sense  but  $C^+(k),C^-(k)$  are decaying  if 
  the initial conditions  for $C$  are smooth. 
  It follows  from   (\ref{eq:2.9}) ,  (\ref{eq:2.12})  that
  \begin{equation}														\label{eq:2.14}
  <C,\Phi>=\int\limits_{\R^2}\big(\overline{C^+}(k)\alpha_k^+-\overline{C^-}(k)\alpha_{-k}^-\big)dk.
  \end{equation}
  It will be   convenient to split  $f_k^+(x_0,x)$  and $f_{-k}^-(x_0,x)$  onto two  parts
  \begin {equation}														\label{eq:2.15}
  f_k^{++}=f_k^+\theta(\eta_\rho),\ \  f_k^{+-}=f_k^+(1-\theta(\eta_\rho)),
  \end{equation}
  where  $\theta(\eta_\rho)=1$, for $\eta_\rho>0,\ \theta(\eta_\rho)=0$ for  $\eta_\rho<0$.      Analogously
  \begin{equation}														\label{eq:2.16}
  f_{-k}^{-+}=f_{-k}^-\theta(\eta_\rho),\ \ f_{-k}^{--}=f_{-k}^-(1-\theta(\eta_\rho)).
  \end{equation}
  Note  that $\overline{f_k^{++}}=f_{-k}^-\theta(\eta_\rho)=f_{-k}^{-+},\ \overline{f_k^{+-}}=f_{-k}^-(1-\theta(\eta_\rho)=f_{-k}^{--}$.
 
 {\bf Remark 2.1}.
 We shall describe  the behavior of  null-bicharacteristics starting  at  $x_0=0$  
 and corresponding to
   $f_k^{++},f_k^{+-},f_{-k}^{-+},f_{-k}^{--}$ that will show  a huge
 difference   between   $f_k^{++},f_{-k}^{--}$  and $f_{k}^{+-},f_{-k}^{-+}$.
 
 The Hamiltonian  corresponding to  $f_k^+$  is
 $\lambda_0^-=-\frac{A}{\rho}\eta_\rho -\frac{B}{\rho^2}\eta_\varphi-\sqrt{\eta_\rho^2+\frac{\eta_\varphi^2}{d^2}}$    (cf.  (\ref{eq:2.3})).
 
 Denote by
 $\gamma^{++}$   the null-bicharacteristic with  the initial data  $(\rho,\varphi,\lambda_0^-,\eta_\rho,\eta_\varphi)$  for  $\eta_\rho>0,\rho>|A|$,
 and  by  $\gamma^{+-}$  the null-bicharacteristic  with  the same  initial data   except  $\eta_\rho<0$.
 
 It is  not difficult  to show  (cf. [E8])   that  $\gamma^{+-}$  crosses  the event horizon  $\{\rho=|A|\}$  when  $x_0\rw +\infty$
 and $\gamma^{++}$  approaches  spiraling  the event  horizon   when  $x_0\rw -\infty$.   One can   say  that  $\gamma^{++}$
 ``emerges" from the  black  hole   when  $x_0$  increases.
 
 Analogously,  the Hamiltonian   for  $f_{-k}^-$   is
  $\lambda_0^+=-\frac{A}{\rho}\eta_\rho -\frac{B}{\rho^2}\eta_\varphi+\sqrt{\eta_\rho^2+\frac{\eta_\varphi^2}{d^2}}$    (cf.  (\ref{eq:2.4})).
 
 If  $\gamma^{-+} (\gamma^{--})$  are  the  null-bicharacteristics  starting  at  $(\rho,\varphi,\lambda_0^+,\eta_\rho,\eta_\varphi)$  where
 $\eta_\rho>0$  for   $\gamma^{-+}$  and  $\eta_\rho<0$   for  $\gamma^{--}$,  then  $\gamma^{-+}$  crosses  the event  horizon   
 when  $x_0\rw +\infty$
and $\gamma^{--}$  approaches  spiraling the event horizon   when  $x_0\rw -\infty$  (cf.  [E8]).

Therefore the  null-bicharacteristics  corresponding  to  $f_k^{++}(x)$  and  $f_k^{--}(k)$,  and   the null-characteristics   corresponding  to  
$f_k^{+-}(x)$   and  $f_k^{-+}(x)$   have  drastically  different behavior.

 Using (\ref{eq:2.15}),  (\ref{eq:2.16})  we can  rewrite  (\ref{eq:2.12})  in  the following  form
 \begin{multline}														\label{eq:2.17}
 C=\int\limits_{\R^2}\big(C^{++}(k)f_k^{++}+C^{+-}(k)f_k^{+-}
 \\
 -C^{-+}(k)f_{-k}^{-+}-C^{--}(k)f_{-k}^{--}\big)dk,
 \end{multline}
 where 
 $$
 C^{++}=C^+\theta(\eta_\rho),C^{+-}=C^+(1-\theta(\eta_\rho)),C^{-+}=C^-\theta(\eta_\rho),C^{--}=C^-(1-\theta(\eta_\rho)).
 $$
 Denote
 \begin{equation}														\label{eq:2.18}
 C_+(\rho,\varphi)=\int_{\eta_\rho>0}C^{++}(k)f_k^{++}(0,x)dk-\int\limits_{\eta_\rho<0}C^{--}(k)f_{-k}^{--}(0,x)dk.
 \end{equation}

 Changing  $\eta_\rho$  to  $-\eta_\rho$  in the second  integral  in  (\ref{eq:2.18})  
 we get
 $$
 C_+(\rho,\varphi)=\int\limits_0^\infty\sum_{m=-\infty}^\infty C_+^{(1)}(k)
 \frac{e^{i\rho\eta_\rho+im\varphi}}{\sqrt \rho\sqrt{\eta_\rho^2+\frac{m^2}{d^2}}}d\eta_\rho.
 $$
 Therefore  $\sqrt \rho C_+(\rho,\varphi)$ admits  an analytic  continuation in $\rho$ to the  half-plane  $\Im\rho>0$.
   Analogously
 \begin{multline}														\label{eq:2.19}
 C_-(\rho,\varphi)=\int\limits_{\eta_\rho<0}C^{+-}(k)f_k^{+-}(0,x)dk-\int\limits_{\eta_\rho>0}C^{-+}(k)f_{-k}^{-+}(0,x)dk
 \\
 =
 \int\limits_{-\infty}^0\sum_{m=-\infty}^\infty C_-^{(1)}(k)
 \frac{e^{i\rho\eta_\rho+im\varphi}}{\sqrt \rho\sqrt{\eta_\rho^2+\frac{m^2}{d^2}}}d\eta_\rho.
 \end{multline}
 Thus
 $\sqrt \rho C_-(\rho,\varphi)$  has  an analytic  continuation  in  $\rho$  to the half-plane  $\Im\rho<0$.  Note that
 \begin{equation}														\label{eq:2.20}
 C(0,\rho,\varphi)=C_++C_-.
 \end{equation}
 Hence   $\sqrt \rho C=\sqrt \rho C_+ +\sqrt \rho C_-$.
 Therefore,  by the well-known formula   (cf.,  for example, [9]),
 \begin{equation}														\label{eq:2.21}
 C_\pm    =\frac{\pm i}{2\pi }\int\limits_{-\infty}^\infty
\sqrt \frac{\rho'}{\rho} \, \,\frac{C(0,\rho',\varphi)d\rho'}{\rho\pm i0-\rho'}
 \end{equation}
 Using  (\ref{eq:2.15}),  (\ref{eq:2.16})  we can  rewrite  (\ref{eq:2.14})  in the form
 \begin{equation}														\label{eq:2.22}
 <C,\Phi>=\int\limits_{\R^2}\big(\overline{C^{++}} \alpha_k^{++} +
 \overline{C^{+-}}\alpha_k^{+-}-\overline{C^{-+}}\alpha_{-k}^{-+}-\overline{C^{--}}\alpha_{-k}^{--}\big)dk,
 \end{equation}
 where
 \begin{multline}														\label{eq:2.23}
 \alpha_k^{++}=<f_k^{++},\Phi>,\ \alpha_k^{+-}=<f_k^{+-},\Phi>,
 \
  \alpha_{-k}^{-+}=<f{_k}^{-+},\Phi>,
  \\
  \alpha_{-k}^{--}=<f_{-k}^{--},\Phi>,
  \ \ \
 \mbox{and}\ \ \ (\alpha_k^{++})^*= \alpha_{-k}^{-+}, \ \ (\alpha_k^{+-})^*= \alpha_{-k}^{--}.
 \end{multline}
 
 We shall define now the vacuum state $|\Psi\rangle$.  
 In  the  case of  Minkowski metric  the vacuum  space  is defined  by the conditions 
 $\alpha_k^{++}|0\rangle =0,\alpha_k^{+-}|0\rangle=0$  for  all $k$.
 It was emphasized  by Unruh [22]  and Jacobson [15]
 the need to modify  the definition of the vacuum states in different situations.  We shall define,  similarly to [23],  [24]),  
 the vacuum state $|\Psi\rangle$  by the requirements
 \begin{align}															\label{eq:2.24}
 &\alpha_k^{++}|\Psi\rangle=0\ \ \mbox{for all}\ \ k=(\eta_\rho,\eta_\varphi) \ \ \mbox{such that}\ \   \eta_\rho>0,
 \\
 \nonumber
 &\alpha_{-k}^{--}|\Psi\rangle=0\ \ \mbox{for all}\ \ k=(\eta_\rho,\eta_\varphi)\ \ \mbox{such that}\ \   \eta_\rho<0.
 \end{align}
 Note that
 $(\alpha_k^{+-})^*=\alpha_{-k}^{--}$. 
 
 It follows  from  (\ref{eq:2.22}),  (\ref{eq:2.24})   that
 \begin{equation}														\label{eq:2.25}
 <C,\Phi>|\Psi\rangle =\int\limits_{\R^2}\big(\overline{C^{+-}}(k)\alpha_k^{+-}-\overline{C^{-+}}(k)\alpha_{-k}^{-+}\big)dk  |\Psi\rangle.
 \end{equation}
 Let  
 \begin{equation}														\label{eq:2.26}
 N(C)=< C,\Phi>^*<C,\Phi>
 \end{equation}
 be the number of particle operator (cf. [14],  [15]).
 
 The expectation  value  of the number  operator 
 \begin{equation}														\label{eq:2.27}
 \langle\Psi|N(C)|\Psi\rangle
 \end{equation}
 is the average  number  of particles created by the wave  packet $C$.
 
 \begin{theorem}                                   									\label{theo:2.1}
 The average number  of particles created by the  wave packet $C$  is given by the formula
 \begin{equation}														\label{eq:2.28}
 \langle \Psi| N(C)|\Psi\rangle=-<C_-,C_->,
 \end{equation}
 where $C_-$  is given  by  (\ref{eq:2.19}).
 \end{theorem}
 {\bf Proof:}
 We have from (\ref{eq:2.24})  that 
 \begin{equation}														\label{eq:2.29}
 \langle\Psi|(\alpha_k^{++})^*=0,\ \ \langle\Psi|(\alpha_{-k}^{--})^*=\langle\Psi|\alpha_k^{+-}=0.
 \end{equation}
 Therefore
 \begin{equation}														\label{eq:2.30}
 \langle\Psi|<C,\Phi>^*=\langle\Psi|\int\limits_{\R^2}\big(C^{+-}(\alpha_k^{+-})^*-C^{-+}(\alpha_{-k}^{-+})^*\big)dk.
 \end{equation}
 Hence,  combining  (\ref{eq:2.30})  and  (\ref{eq:2.25})  we get
 \begin{multline}														
 \nonumber
  \langle\Psi|<C,\Phi>^*<C,\Phi>|\Psi\rangle
 =\langle\Psi|\int\limits_{\R^2}\big(C^{+-}(\alpha_k^{+-})^*-C^{-+}(\alpha_{-k}^{-+})^*\big)dk
 \\
\cdot \int\limits_{\R^2}\big(\overline{C^{+-}}(k')\alpha_{k'}^{+-}-\overline{C^{-+}}(k')\alpha_{-k'}^{-+})\big)dk'|\Psi\rangle.
 \end{multline}
 Therefore
 \begin{equation}														\label{eq:2.31}
  \langle\Psi|N(C)|\Psi\rangle
   =\int\limits_{\R^2}\big( -|C^{+-}(k)|^2+|C^{-+}(k)|^2\big)dk.
 \end{equation}
We  used  that  
\begin{multline}
\nonumber
(\alpha_k^{+-})^*\alpha_{k'}^{+-}=
\alpha_{k'}^{+-}(\alpha_k^{+-})^*-I\delta(k-k'),
\\
\ \
\alpha_k^{++}(\alpha_{k'}^{++})^*-(\alpha_{k'}^{++})^*\alpha_k^{++}=I\delta(k-k'),
\ \
 \alpha_{-k}^{-+}=(\alpha_k^{++})^*.\ \ \ \
\end{multline}
It follows  from  (\ref{eq:2.19})  that
\begin{equation}														\label{eq:2.32}
<C_-,C_->=
\int\limits_{\R^2}\big(|C^{+-}(k)|^2-|C^{-+}(k)|^2\big)dk
\end{equation}
Therefore
\begin{equation}														\label{eq:2.33}
\langle\Psi|N(C)|\Psi\rangle=-<C_-,C_->,
\end{equation}
where $C_-$  is given by  (\ref{eq:2.19}).

\section{Hawking radiation in the case of rotating black hole}
\init

The Hamiltonian   of  (\ref{eq:1.1})  has  the following  form  in polar coordinates  $(\rho,\varphi)$
$
H(\rho,\varphi,\xi_0,\xi_\rho,m)=\big(\xi_0+\frac{A}{\rho}\eta_\rho  +\frac{B}{\rho^2}m\big)^2-\xi_\rho^2-\frac{m^2}{\rho^2}=0,
$
where  $(\xi_0,\eta_\rho,m)$  are  dual variables  to  $(x_0,\rho,\varphi)$.

Let  $S=-\eta_0x_0+S_1(\rho)+m\varphi$  be the solution of the eikonal equation 
\begin{equation} 														\label{eq:3.1}
\Big(-\eta_0+\frac{A}{\rho} S_{1\rho} +\frac{B}{\rho^2}m\Big)^2-S_{1\rho}^2-\frac{m^2}{\rho^2}=0.
\end{equation}
We are looking for the solution  of (\ref{eq:3.1})  such that $S_{1\rho}\rw +\infty$  when  $\rho\rw |A|$.
Solving  the  quadratic equation  (\ref{eq:3.1})  we get
\begin{equation}														\label{eq:3.2}
S_{1\rho}=\frac{\frac{A}{\rho}\big(\eta_0-\frac{Bm}{\rho^2}\big)\pm
\sqrt{\frac{A^2}{\rho^2}\big(\eta_0-\frac{Bm}{\rho^2}\big)^2-\big(\frac{A^2}{\rho^2}-1\big)\big[\big(\eta_0-\frac{B}{\rho^2}m\big)^2
}
-\frac{m^2}{\rho^2}\big]
}
{\frac{A^2}{\rho^2}-1}.
\end{equation}
Let
\begin{equation}														\label{eq:3.3}
\xi_0=\eta_0-\frac{B}{|A|^2}m.
\end{equation}
Since we are looking  for $S_{1\rho}\rw +\infty$  when   $\rho -|A|\rw 0$  we have
\begin{equation}														\label{eq:3.4}
S_{1\rho}=\frac{\xi_0|A|}{\rho-|A|}+O(\rho-|A|).
\end{equation}
Thus  $S_1=\xi_0|A|\ln(\rho-|A|)+O(\rho-|A|)$.

 We define  a wave  packet  $C_0(x_0,\rho,\varphi)$  as the exact  solution of the wave  equation  (\ref{eq:1.1})  
with the following   initial  conditions
\begin{equation}															\label{eq:3.5}
C_0\big|_{x_0=0}=\theta(\rho-|A|)\frac{1}{\sqrt \rho}(\rho-|A|)^{\e} e^{-a(\rho-|A|)}\ e^{i\xi_0|A|\ln(\rho-|A|)+im\varphi},
\end{equation}
\begin{equation} 															\label{eq:3.6}
\frac{\partial C_0}{\partial x_0}\Big|_{x_0=0}=
i\beta \theta(\rho-|A|)\frac{1}{\sqrt \rho}(\rho-|A||)^{\e}e^{-a(\rho-|A|)}\ e^{i\xi_0|A|\ln(\rho-|A|)+im\varphi},
\end{equation}
where  $a>0,\e>0$,
\begin{align}																\label{eq:3.7}
\beta  &= -\frac{A}{\rho}\frac{\xi_0 |A|}{\rho-|A|}-\frac{B}{\rho^2}m-\frac{\xi_0|A|}{\rho-|A|}=
\big(\frac{|A|}{\rho}-1\big)\frac{\xi_0|A|}{\rho-|A|}-\frac{B}{\rho^2}m
\\
\nonumber
&=-\frac{\xi_0|A|}{\rho}-\frac{Bm}{\rho^2}=-\eta_0+  O(\rho-|A|).
\end{align}
We used in (\ref{eq:3.7})  that  $A<0$  and  $\xi_0=\eta_0-\frac{Bm}{|A|^2}$.

The  convenience of this choice  of  the initial  conditions  (\ref{eq:3.5}),  (\ref{eq:3.6})   will be  clear  later.

We shall compute  the  KG norm of  $C_0$.

Let,  as  in  (\ref{eq:2.6}),  $\{v_1,v_2\}=\sum_{j=0}^2\big(\overline v_1\frac{\partial v_2}{\partial x_j}-
\frac{\partial\overline v_1}{\partial x_j}v_2\big)$.
Note that  if  $C_0=h(\rho,\varphi) C_{01}$,  where $h(\rho,\varphi)=\frac{f(\rho)}{\sqrt \rho}$  is real-valued,  then
$\{C_0,C_0\}=h^2(\rho,\varphi)\{C_{01},C_{01}\}$.  Therefore
$$
\{C_0,C_0\}=  \frac
{\theta(\rho-|A|)f^2(\rho)(-2i\xi_0|A|)}
{\rho(\rho-|A|)}.
$$
Thus
\begin{equation}												\label{eq:3.8}
<C_0,C_0>=\int\limits_{|A|}^\infty\int\limits_0^{2\pi}\frac{2\xi_0|A|}{\rho-|A|}(\rho-|A|)^{2\e}e^{-2a|\rho-A|}d\rho d\varphi
=\frac{4\pi\xi_0|A|\Gamma(2\e)}{(2a)^{2\e}},
\end{equation}
since  $f(\rho)=(\rho-|A|)^\e e^{-a(\rho-|A|)}$.

We are going to compute $\langle \Psi|N(C_0)|\Psi\rangle$  (cf.  (\ref{eq:2.26})), 
 i.e.  the average  number  of particles  created  by the wave packet  $C_0$.  

It follows  from   (\ref{eq:2.32})   that
\begin{equation}															\label{eq:3.9}
\langle \Psi | N(C_0)|\Psi\rangle =\int\limits_{\R^2}(|C_0^{-+}(k)|^2-|C_0^{+-}(k)|^2)dk,
\end{equation}
where
\begin{equation}
\nonumber															
C_0^{+-}(k)=<f_k^{+-},C_0>,\ \ C_0^{-+}(k)=<f_{-k}^{-+},C_0>.
\end{equation}
We have  (cf.  (\ref{eq:2.1}),  (\ref{eq:2.2}),  (\ref{eq:3.5}),  (\ref{eq:3.6}))
\begin{equation}                         												\label{eq:3.10}
C_0^{+-}(k)=C_1^{+-}(k)+C_2^{+-}(k),
\end{equation}
where
\begin{multline}															\label{eq:3.11}
C_1^{+-}(k)=
\int\limits_0^{2\pi}\int\limits_0^{\infty}\frac{\big(\eta_\rho^2+\frac{m'^2}{d^2}\big)^{\frac{1}{4}} }
{\sqrt 2\, 2\pi\sqrt\rho}
e^{-i\rho\eta_\rho-im'\varphi}
\\
\cdot \frac
{\theta(\rho-|A|)(\rho-|A|)^{\e}}
{\sqrt \rho}
 e^{-a(\rho-|A|)+i\xi_0|A|\ln(\rho-|A|) +im\varphi}\rho d\rho  d\varphi,
\end{multline}
$k=(\eta_\rho,m'),\eta_\rho<0$.  

Analogously,
\begin{multline}															\label{eq:3.12}
C_2^{+-}(k)=
\int\limits_0^{2\pi}\int\limits_0^{\infty}\frac{e^{-i\rho\eta_\rho-im'\varphi}
}
{\sqrt 2\, 2\pi\sqrt\rho   \big(\eta_\rho^2+\frac{m'^2}{d^2}\big)^{\frac{1}{4}} }
\cdot \frac
{\theta(\rho-|A|)(\rho-|A|)^{\e}}
{\sqrt \rho}
\\
\cdot
\Big(
\frac{\xi_0|A|}{\rho-|A|}-\frac{A}{\rho}\Big(\frac{-i\e}{\rho-|A|}+ia\Big)\Big)
 e^{-a(\rho-|A|)+i\xi_0|A|\ln(\rho-|A|) +im\varphi}\rho d\rho  d\varphi,
\end{multline}
where  while computing the KG  inner product  we used that  (cf.  (\ref{eq:3.9}))
$$
\beta +\frac{A}{\rho}\frac{\xi_0|A|}{\rho-|A|}+ \frac{B}{\rho^2}m=-\frac{\xi_0|A|}{\rho-|A|}.
$$
Since  $\int_0^{2\pi}e^{-im'\varphi+im\varphi}d\varphi  
=2\pi\delta_{mm'}$  where $\delta_{mm'}=1$  when  $m=m'$  and 
$\delta_{mm'}=0$  when  $m\neq m'$,  we get
\begin{equation}														\label{eq:3.13}
C_1^{+-}=\delta_{mm'}\int\limits_0^\infty\frac{1}{\sqrt 2}\Big(\eta_\rho^2+\frac{m^2}{d^2}\Big)^{\frac{1}{4}}
\theta(\rho-|A|)e^{-a(\rho-|A|)}
e^{(i\xi_0|A|+\e)\ln(\rho-|A|)-i\rho\eta_\rho}d\rho.
\end{equation}   
Change variable  $\rho-|A|=x  $  in  (\ref{eq:3.13})  
and  perform the  Fourier transform in $x$.  Using  the well-known  formula  
\begin{equation}														\label{eq:3.14}
F(x_+^\lambda e^{-ax})=\frac{\Gamma(\lambda+1)e^{-i(\lambda+1)\frac{\pi}{2}}}{(\eta_1-ia)^{\lambda+1}},
\end{equation}
 (see,
for  example,  formula   (11.10)  in [E7]),
we get,  having  $\lambda=i\xi_0|A|+\e,$
\begin{equation}														\label{eq:3.15}
C_1^{+-}=\frac{\delta_{mm'}}{\sqrt 2}
\frac{\big(\eta^2_\rho+\frac{m^2}{d^2}\big)^{\frac{1}{4}}e^{-i|A|\eta_\rho} \Gamma(i\xi_0|A|+\e+1)
e^{-i(i\xi_0|A|+\e+1)\frac{\pi}{2}}}
{(\eta_\rho-ia)^{i\xi_0|A|+\e+1}}.
\end{equation}
Analogously,
\begin{multline}														\label{eq:3.16}
C_2^{+-}=\frac{\delta_{mm'}(\xi_0|A|-i\e)}{\sqrt 2}
\frac{\big(\eta^2_\rho+\frac{m^2}{d^2}\big)^{-\frac{1}{4}}e^{-i|A|\eta_\rho} \Gamma(i\xi_0|A|+\e)
e^{-i(i\xi_0|A|+\e)\frac{\pi}{2}}}
{(\eta_\rho-ia)^{i\xi_0|A|+\e}}
\\
+\frac{\delta_{mm'}\cdot O\big(\frac{1}{|\eta_\rho-ia|^{\e+1}}\big)}{\sqrt 2\big(\eta_\rho^2+\frac{m^2}{d^2}\big)^{\frac{1}{4}}}.
\end{multline}

Noting that  $\eta_\rho<0, a>0$  we  get
$$
\ln(\eta_\rho-ia)=\ln|\eta_\rho-ia|+i\big(-\pi +\sin^{-1}\frac{a}{\sqrt{\eta_\rho^2+a^2}}\big).
$$
Also
$\Gamma(\lambda+1)=\lambda \Gamma(\lambda)$.  Therefore
\begin{multline}															\label{eq:3.17}
C_1^{+-}\overline C_2^{+-}=\delta_{mm'}\frac{\xi_0|A|+i\e}{2}(i\xi_0|A|+\e) |\Gamma(i\xi_0|A|+\e)|^2
e^{-i\frac{\pi}{2}+\pi \xi_0|A|}
\\
\cdot \frac
{e^{-\big(2\pi-2\sin^{-1}\frac{a}{\sqrt{\eta_\rho^2+a^2}}\big)\,\xi_0|A|}}
{|\eta_\rho-ia|^{2\e}(\eta_\rho-ia)}+O\Big(\frac{1}{|\eta_\rho-ia|^{2\e+2}}\Big).
\end{multline}
Note that
\begin{equation} 															\label{eq:3.18}
\Gamma(i\xi_0|A|+\e)=\int\limits_0^\infty e^{(i\xi_0|A|+\e-1)\ln x-x}dx.
\end{equation}
Using the Cauchy theorem we  can  replace 
the integration  over  real semiaxis  by the integration  over  the imaginary  semiaxis:
\begin{equation}															\label{eq:3.19}
\Gamma(i\xi_0|A|+\e)=i\int\limits_0^\infty e^{(i\xi_0|A|+\e-1)(\ln y+i\frac{\pi}{2})-iy}dy=
e^{-\frac{\pi}{2}\xi_0|A|}\Gamma_1(\xi_0|A|),
\end{equation}
where 
$$
\Gamma_1(\xi_0|A|)=i\int\limits_0^\infty e^{(i\xi_0|A|+\e-1)\ln y+i(\e-1)\frac{\pi}{2}-iy}dy.
$$
Therefore
\begin{multline}															\label{eq:3.20}
\
\\
C_1^{+-}\overline C_2^{+-}
=\delta_{mm'}\frac{(\xi_0|A|)^2+\e^2}{2}
 |\Gamma_1|^2               
{
e^{-\big(2\pi-2\sin^{-1}\frac{a}{\sqrt{\eta_\rho^2+a^2}}\big)\xi_0|A|}
}
\\
\cdot  \frac{(\eta_\rho+ia)}
{|\eta_\rho-ia|^{2\e+2}}
+O\big(\frac{1}{|\eta_\rho-ia|^{2\e+2}}\big),
\ \ \ \eta_\rho<0.
\end{multline}
Consider  now  $C^{-+}(k)=<f_{-k}^-\theta(\eta_\rho),C_0>$.
When we change  $\eta_\rho$  to  $-\eta_\rho$  then  the only difference  with  (\ref{eq:3.10}),  (\ref{eq:3.11})  is
that  $\big(\eta_\rho^2+\frac{m^2}{d^2}\big)^{\frac{1}{4}}$ in  (\ref{eq:3.11})  is replaced by 
$-\big(\eta_\rho^2+\frac{m^2}{d^2}\big)^{\frac{1}{4}}$.

Therefore  
$$
C^{-+}=-C_1^{+-}+C_2^{+-}.
$$
We have 
\begin{align} 													\nonumber
&|C^{+-}|^2=|C_1^{+-}+C_2^{+-}|^2=|C_1^{+-}|^2+C_1^{+-}\overline C_2^{+-} +C_2^{+-}\overline C_1^{+-}
+|C_2^{+-}|^2,
\\
\nonumber
&|C^{-+}|^2=|C_1^{+-}|^2-C_1^{+-}\overline C_2^{+-}-C_2^{+-}\overline C_1^{+-}+|C_2^{+-}|^2.
\end{align}
Therefore
\begin{equation}															\label{eq:3.21}
\langle \Psi |N(C_0)|\Psi\rangle =\sum_{m'=-\infty}^\infty\int\limits_{-\infty}^0( |C^{-+}|^2-|C^{+-}|^2)d\eta_\rho
=-
\sum_{m'=-\infty}^\infty\int\limits_{-\infty}^0 4\Re (C_1^{+-}\overline C_2^{+-})d\eta_\rho.
\end{equation}
It follows  from (\ref{eq:3.20})  that
\begin{multline}																\label{eq:3.22}
\langle \Psi |N(C_0)|\Psi\rangle =
2e^{-2\pi |\xi_0|A|}
(( |\xi_0|A|)^2+\e^2)
\, |\Gamma_1|^2
\int\limits_{-\infty}^0
\frac
{
|\eta_\rho|
e^{2\xi_0|A|\sin^{-1}\frac{a}{\sqrt{\eta_\rho^2+a^2}}     
}
}
{(\eta_\rho^2+a^2)^{\e+1}}
d\eta_\rho
\\
+\int\limits_{-\infty}^0 O\big(\frac{1}{|\eta_\rho-ia|^{2\e+2}}\big) d\eta_\rho.
\end{multline}
Making the change of variable $\eta_\rho\rw a\eta_\rho$,  we get
\begin{multline}																\label{eq:3.23}
\langle \Psi |N(C_0)|\Psi\rangle =
2e^{-2\pi |\xi_0|A|} 
(( |\xi_0|A|)^2+\e^2)
\, a^{-2\e}|\Gamma_1|^2
\int\limits_{-\infty}^0
\frac
{
|\eta_\rho|
e^{2\xi_0|A|\sin^{-1}\frac{1}{\sqrt{\eta_\rho^2+1}}     
}
}
{(\eta_\rho^2+1)^{\e+1}}
d\eta_\rho
\\
+
O(a^{-2\e-1}).
\end{multline}

We now  normalize  $C_0$  replacing it  by $C_n=\frac{C_0}{<C_0,C_0>^{\frac{1}{2}}}$.

We have  $N(C_n)=\frac{N(C_0)}{<C_0,C_0>}$.

Noting  that  $<C_0,C_0>=\frac
{4\pi \xi_0|A|\Gamma(\e)}
{(2a)^{\frac{1}{2}}
}
$  (see (\ref{eq:3.8}))   we get  from  (\ref{eq:3.23})  that
\begin{multline}																	\label{eq:3.24}
\lim_{a\rw\infty}\langle \Psi |N(C_n)|\Psi\rangle  =
\frac{2^{2\e}e^{-2\pi\xi_0|A|}|\Gamma_1|^2}{2\pi \Gamma(\e)}
\\
\cdot \frac{(\xi_0|A|)^2+\e^2}{\xi_0|A|}\int\limits_{-\infty}^0\frac{|\eta_\rho|}{(\eta_\rho^2+1)^{\e+1}}\ 
e^{2\xi_0|A|\sin^{-1}\frac{1}{\sqrt{\eta_\rho^2+1}}}
d\eta_\rho.
\end{multline}

\begin{theorem}																	\label{theo:3.1}
Let  $\xi_0=\eta_0-\frac{Bm}{|A|^2}$  and let $C_0(x_0,\rho,\varphi)$  be the solution  of  (\ref{eq:1.1})
with the initial conditions (\ref{eq:3.5}),  (\ref{eq:3.6}).  Then  the  average  number of particles created  by the  
normalized wave packet  
$C_n(x_0,\rho,\varphi)$ is  given  by (\ref{eq:3.24})   when  $a\rw\infty$.
\end{theorem}

{\bf Remark 3.1} 
 K.Fredenhagen  and R.Haag  (see [11])  use the  limit   when  the time
$T$  tends  to $-\infty$ to find the Hawking  radiation  in a spherically symmetric case.  Note  that  when $T\rw -\infty$ the wave  packet  becomes  closer and closer  to the black hole.  In our approach the time is fixed  and  in   $(\rho-|A|)^\e e^{-a(\rho-|A|)}$  the parameter $a$
characterizes the closeness  to the black hole.  Thus,  as $a\rw +\infty$   we get  the  value  of the  Hawking radiation.

\section{Hawking radiation  for analogue black holes  with variable $A$ and $B$}
\init

In this section we extend the results of \S 3  to the case when 
$A$  and $B$  depend on $(\rho,\varphi), \rho>0, \varphi\in [0,2\pi]$ 
(see  [6],  [7],  [18],  [19], [23], [27],  [28],  where physically relevant examples of such acoustic metrics are studied).

\subsection{The case  of $A<0$  constant and  $|B(\varphi)|>0$  periodic}

Consider first a more simple case when $A<0$  is a constant  and $B(\varphi)$  is a periodic function  of $\varphi,\ |B(\varphi)|>0$.  
In this case  $\rho=\sqrt{A^2+B^2(\varphi)}$  is an ergosphere and  $\{\rho<|A|\}$  is a black hole. 

The eikonal $S=-x_0\eta_0+S_1(\rho,\varphi)$  is the solution of the equation
\begin{equation}																		\label{eq:4.1}
\Big(-\eta_0+\frac{A}{\rho}\frac{\partial S_1}{\partial \rho}+\frac{B(\varphi)}{\rho^2}\frac{\partial S_1}{\partial \varphi}\Big)^2
-\Big(\frac{\partial S_1}{\partial\rho}\Big)^2-\frac{1}{\rho^2}\Big(\frac{\partial S_1}{\partial \varphi}\Big)^2=0,
 \ \rho>|A|,\varphi\in[0,2\pi].
\end{equation}

Solving the quadratic equation  (\ref{eq:4.1})  we get (cf.  (\ref{eq:3.2}))
\begin{equation}														             			\label{eq:4.2}
S_{1\rho}=\frac{
\frac{A}{\rho}\big(\eta_0-\frac{B(\varphi)}{\rho^2}\frac{\partial S_1}{\partial\varphi}\big)
\pm\sqrt{
\big(\eta_0-\frac{B(\varphi)}{\rho^2}\frac{\partial S_1}{\partial\varphi}\big)^2+\big(\frac{A^2}{\rho^2}-1\big)
\frac{\big(\frac{\partial S_1}{\partial\varphi}\big)^2}
{\rho^2}
}
}
{\frac{A^2}{\rho^2}-1}
\end{equation}
 We shall find  a simple  approximation   for the eikonal 
  near  $\rho=|A|.$   

Assuming  that  $S_{1\varphi}(\rho,\varphi)$  is  continuous  at  $\rho=|A|$,  we get from  (\ref{eq:4.2})
$$
S_{1\rho}=\frac{\big(\eta_0-\frac{B(\varphi)}{|A|^2}S_{1\varphi}(\rho,\varphi)\big)|A|}{\rho-|A|}+O(1).
$$
Denote by  $S_2(\rho,\varphi)$  the solution  of the first order  partial differential  equation
\begin{equation}																	\label{eq:4.3}
(\rho-|A|)S_{2\rho}(\rho,\varphi)-\eta_0|A|+  \frac{B(\varphi)}{|A|}S_{2\varphi}(\rho,\varphi)=0.
\end{equation}
There is an alternative  way  to obtain  the approximation  of $S_1(\rho,\varphi)$  by  $S_2(\rho,\varphi)$:  Consider  the eikonal  equation
$$
-\eta_0+\frac{A}{\rho}S_{1\rho}+\frac{B(\varphi)}{\rho^2} S_{1\varphi}=-\sqrt{S_{1\rho}^2+\frac{1}{\rho^2}S_{1\varphi}^2}.
$$
We are  looking  for  $S_1$  such  that  $S_1\rw +\infty$  when  $\rho\rw |A|$. 
 Then $\sqrt{S_{1\rho}^2+\frac{1}{\rho^2}S_{1\varphi}^2}
 =S_{1\rho}\big(1+\frac{\S_{1\varphi}^2}{\rho^2S_{1\rho}^2}\big)^{\frac{1}{2}}$,  where $\frac{\S_{1\varphi}^2}{\rho^2S_{1\rho}^2}$
is small.
Replacing $\sqrt{S_{1\rho}^2+\frac{1}{\rho^2}S_{1\varphi}^2}$  by  $S_{1\rho}$   we get  the linear  equation  
$$
-\eta_0+\frac{\rho -|A|}{\rho}S_{1\rho}+\frac{B(\varphi)}{\rho^2} S_{1\varphi}=0,
$$
which becomes  (\ref{eq:4.3})   when  $\rho$  is replaced  by $ |A|$.

Equation  (\ref{eq:4.3})  can be solved explicitly  and we take  $S_2(\rho,\varphi)$  as  the approximation  of  $S_1(\rho,\varphi)$.
We have 
$$
S_2(\rho,\varphi)=\eta_0|A|\ln(\rho-|A|) +S_3(\rho,\varphi),
$$
where $S_3(\rho,\varphi)$  is  the solution  of the  homogeneous  equation:
$$
(\rho-|A|)S_{3\rho}(\rho,\varphi)+\frac{B(\varphi)}{|A|}S_{3\varphi}=0.
$$
Consider  the characteristic equation
$$
\frac {d\rho}{\rho -|A|}-\frac{|A|d\varphi}{B(\varphi)}=0.
$$
Then  
$$
S_3(\rho,\varphi)=g\Big(\ln(\rho-|A|)-\int\limits_0^\varphi\frac{|A|d\varphi'}{B(\varphi')}\Big)
$$
is the  general  solution of  a  homogeneous equation  for arbitrary  $g(t)$.  
We take $g(t)=a_0t$,  where  $a_0$  is such  that  $S_3(\rho,\varphi+2\pi)=S_3(\rho,\varphi)+2\pi m, \  m\in\Z$. 
 Thus  $a_0$  satisfies  the equation     
$a_0\int_0^{2\pi}\frac{d\varphi}{B(\varphi)}|A|=2\pi m$,  i.e.
\begin{equation}																	\label{eq:4.4}
a_0=\frac{2\pi m}{|A|\int_0^{2\pi}B^{-1}(\varphi) d\varphi}.
\end{equation}
Therefore
$$
S_3(\rho,\varphi)=a_0\ln|\rho-|A||+S_4(\varphi),
$$
where 
\begin{equation}																	\label{eq:4.5}
S_4(\varphi)=-a_0\int\limits_0^\varphi\frac{|A|d\varphi'}{B(\varphi')},
\end{equation}
Finally,
$$
S_2(\rho,\varphi)=(\eta_0|A|+a_0)\ln (\rho-|A|) +S_4(\varphi).
$$

Define the wave packet  $\hat C_0$  as  the solution  of the wave equation 
$\Box_gu=0$  having  the initial conditions (cf. (\ref{eq:3.5}),   (\ref{eq:3.6}))
\begin{align}																			\label{eq:4.6}
&\hat C_0\big|_{x_0=0}=\theta(\rho-|A|) f(\rho)e^{i(\eta_0|A|+a_0)\ln(\rho-|A|)+iS_4(\varphi)},
\\
                                             																\label{eq:4.7}
&\frac{\partial\hat C_0}{\partial x_0}\big|_{x_0=0}=i\hat\beta\theta(\rho-|A|) f(\rho)
e^{i(\eta_0|A|+a_0)\ln(\rho-|A|)+iS_4(\varphi)},
\end{align}
where $a_0$ and $S_4(\varphi)$  are the same as  in (\ref{eq:4.4}),  (\ref{eq:4.5}),
 $ f(\rho)=\frac{1}{\sqrt \rho}(\rho-|A|)^\e e^{-a(\rho-|A|)}$,   and
     (cf.  (\ref{eq:3.7}))
  \begin{align}																		\label{eq:4.8}
  \hat\beta =&-\frac{A}{\rho}\frac{(\eta_0|A|+a_0)}{\rho-|A|}-\frac{B(\varphi)}{\rho^2}\,
  \frac{\partial S_4}{\partial\varphi}-\frac{\eta_0|A|+a_0}{\rho-|A|}
  \\ 
  \nonumber
  =&\big(\frac{|A|}{\rho}-1\big)\frac{\eta_0|A|+a_0}{\rho-|A|}-\frac{B(\varphi)}{\rho^2}(-a_0|A|B^{-1}(\varphi))
  \\
  \nonumber
  =&-\frac{\eta_0|A|+a_0}{\rho}+\frac{a_0|A|}{\rho^2}=-\eta_0+O(\rho-|A|).
\end{align}
The computation  of $\langle\Psi|N(\hat C_0)|\Psi\rangle$  are  the same as in  \S3 with  $\xi_0|A|$  replaced
by $\eta_0|A|+a_0$  and  $e^{im\varphi}$  replaced  by  $e^{iS_4(\varphi)}$.
Let 
\begin{equation}																	\label{eq:4.9}
\hat C_0^{+-}= \hat  C_1^{+-} +\hat C_2^{+-},
\end{equation}
(cf.  (\ref{eq:3.10})).   Then,  as in  (\ref{eq:3.11}))
\begin{multline}															\label{eq:4.10}
\hat C_1^{+-}=
\int\limits_0^{2\pi}\int\limits_0^{\infty}\frac{\big(\eta_\rho^2+\frac{m'^2}{d^2}\big)^{\frac{1}{4}} }
{\sqrt 2\, 2\pi\sqrt\rho}
e^{-i\rho\eta_\rho-im'\varphi}
\\
\cdot \frac
{\theta(\rho-|A|)(\rho-|A|)^{\e}}
{\sqrt \rho}
 e^{-a(\rho-|A|)+i(\eta_0|A|+a_0)\ln(\rho-|A|) +iS_4(\varphi))}\rho d\rho \,d\varphi,            
\end{multline}
Analogously  (cf.  (\ref{eq:3.12}))
\begin{multline}															\label{eq:4.11}
\hat C_2^{+-}=
\int\limits_0^{2\pi}\int\limits_0^{\infty}\frac{e^{-i\rho\eta_\rho-im'\varphi}
}
{\sqrt 2\, 2\pi\sqrt\rho   \big(\eta_\rho^2+\frac{m'^2}{d^2}\big)^{\frac{1}{4}} }
\\
\cdot \frac
{\theta(\rho-|A|)(\rho-|A|)^{\e}}
{\sqrt \rho}
\Big(
\frac{(\eta_0|A|+a_0)}{\rho-|A|}
-\frac{A}{\rho}\Big(\frac{-i\e}{\rho-|A|}+i a\Big)
\Big)
\\
\cdot 
 e^{-a(\rho-|A|)+i(\eta_0|A|+a_0)\ln(\rho-|A|) +iS_4(\varphi)}\rho d\rho d\varphi,
\end{multline}
Let
\begin{equation}														\label{eq:4.12}
\hat\gamma_{m'}=\frac{1}{2\pi}\int\limits_0^{2\pi}e^{-im'\varphi+iS_4(\varphi)}d\varphi.
\end{equation}
Then,  integrating  in  $\rho$ and in $\varphi$  as in  (\ref{eq:3.15}),  (\ref{eq:3.16}),
we get 
\begin{equation}														\label{eq:4.13}
\hat C_1^{+-}=\frac{\hat\gamma_{m'}}{\sqrt 2}
\frac{\big(\eta^2_\rho+\frac{m'^2}{d^2}\big)^{\frac{1}{4}}e^{-i|A|\eta_\rho} \Gamma(i(\eta_0|A|+a_0)+\e+1)
e^{-i(i\eta_0|A|+a_0)+\e+1)\frac{\pi}{2}}}
{(\eta_\rho-ia)^{i(\eta_0|A|+a_0)+\e+1}}.
\end{equation}
Analogously,
\begin{multline}														\label{eq:4.14}
\hat C_2^{+-}=\frac{\hat\gamma_{m'} ((\eta_0|A|+a_0)-i\e)}{\sqrt 2}
\\
\cdot
\frac{\big(\eta^2_\rho+\frac{m'^2}{d^2}\big)^{-\frac{1}{4}}e^{-i|A|\eta_\rho} \Gamma(i(\eta_0|A|+a_0)+\e)
e^{-i((i\eta_0|A|+a_0)+\e)\frac{\pi}{2}}}
{(\eta_\rho-ia)^{i(\eta_0|A|+a_0)+\e}}
\\
+\frac{\gamma_{mm'}}{\sqrt 2\big(\eta_\rho^2+\frac{m^2}{d^2}\big)}
O\Big(\frac{1}{|\eta_\rho-ia|^{\e+1}}\Big).
\end{multline}
Therefore,   integrating in  $\eta_\rho$  and summing in  $m'$,   we get   (cf.   (\ref{eq:3.23})):  
\begin{multline}																\label{eq:4.15}
\langle \Psi |N(\hat C_0)|\Psi\rangle 
\\
=
2e^{-2\pi (\eta_0|A|+a_0)} 
\sum_{m'=-\infty}^\infty|\hat\gamma_{m'}|^2
((\eta_0|A|+a_0)^2+\e^2)\, a^{-2\e}|\Gamma_1(\eta_0|A|+a_0)|^2
\\
\cdot
\int\limits_{-\infty}^0
\frac
{
|\eta_\rho|
e^{2(\eta_0|A|+a_0)\sin^{-1}\frac{1}{\sqrt{\eta_\rho^2+1}}     
}
}
{(\eta_\rho^2+1)^{\e+1}}
d\eta_\rho  +O(a^{-2\e-1}).
\end{multline}

The Parseval's  equality  gives
\begin{equation}																\label{eq:4.16}
\sum_{-m'=-\infty}^\infty  |\hat\gamma_{m'}|^2=\frac{1}{2\pi}\int\limits_0^{2\pi}|e^{iS_4(\varphi)}|^2d\varphi=1.
\end{equation}
If we replace  $\hat C_0$  by  the normalized  wave packet  $\hat C_n=\frac{\hat C_0}{<\hat C_0,\hat C_0>}$  and  take 
 the limit  as  $a\rw\infty$,    we get
\begin{multline}																	\label{eq:4.17}
\lim_{a\rw \infty}\langle \Psi| N(\hat C_n) |\Psi\rangle 
\\
=\frac{2^{2\e}}{2\pi \Gamma(\e)}e^{-2\pi(\eta_0|A|+a_0)}
|\Gamma_1(\eta_0|A|+a_0)|^2
\frac
{(\eta_0|A|+a_0)^2+\e^2}{\eta_0|A|+a_0}
\\
\cdot
\int\limits_{-\infty}^0\frac{|\eta_\rho|}{(\eta_\rho^2+1)^{1+\e}} 
e^{2(\eta_0|A|+a_0)\sin^{-1}\frac{1}{\sqrt{\eta_\rho^2+\e}}}d\eta_\rho.
\end{multline}

We proved  the following theorem:
\begin{theorem}																	\label{theo:4.1}
Let  $a_0$  be the same  as in  (\ref{eq:4.4}).  The average  number of particles  of created  by the  
normalized
wave  packet
$\hat C_n(x_0,\rho,\varphi)$  is given  by  (\ref{eq:4.17})   when  $a\rw\infty$.
\end{theorem}


\subsection{General acoustic metric}

Now  consider  acoustic metrics  of general form,  i.e.  when  $A(\rho,\varphi),B(\rho,\varphi)$  are functions  of  $(\rho,\varphi),
A(\rho,\varphi)<0,|B(\rho,\varphi)|>0$.  The equation of the ergosphere is  $\frac{A^2+B^2}{\rho^2}=1$
and we assume that the ergosphere  is a smooth Jordan curve.  It was shown  in [6], [7]   that there exists a black hole 
$\{\rho<\rho_0(\varphi)\}$  inside  the ergosphere where $\rho=\rho_0(\varphi)$  is a smooth Jordan  curve that is a characteristic  curve
for the spatial   part of the wave   operator  $\Box_g$.  Let
\begin{equation}																			\label{eq:4.16}  
\Big(-\eta_0+\frac{A(\rho,\varphi)}{\rho}S_\rho+\frac{B(\rho,\varphi)}{\rho^2}S_\varphi\Big)^2-S_\rho^2-\frac{1}{\rho^2}S_\varphi^2=0
\end{equation}
be the eikonal  equation  (cf.  (\ref{eq:4.1})).

Make  change of variable
\begin{align}																			\label{eq:4.19}    
&\tilde \rho=\rho-\rho_0(\varphi),
\\
\nonumber
&\varphi=\varphi.
\end{align}

Let  $\tilde S(\tilde\rho,\varphi)=S(\rho,\varphi)$.   The eikonal  equation (\ref{eq:4.16})  has  the following form  in
$(\tilde\rho,\varphi)$ coordinates
\begin{equation}																		\label{eq:4.20}     
\Big(-\eta_0+\Big(\frac{A}{\rho}-\frac{B}{\rho^2}\rho_0'(\varphi)\Big)\tilde S_{\tilde\rho}+\frac{B}{\rho^2}\tilde S_\varphi\Big)^2
-\tilde S_{\tilde\rho}^2-\frac{1}{\rho^2}\big(\tilde S_{\varphi}-\rho_0'(\varphi)\tilde S_{\tilde\rho}\big)^2=0,
\end{equation}
where
$\rho=\rho_0(\varphi)+\tilde\rho,\rho_0'(\varphi)=\frac{d\rho_0(\varphi)}{d\varphi}$.

We shall  rewrite  (\ref{eq:4.20})  as a quadratic equation  in  $\tilde S_{\tilde\rho}$:
\begin{equation}																		\label{eq:4.21}     
\tilde A(\tilde\rho,\varphi)\tilde S_{\tilde\rho}^2 +2\tilde B(\eta_0,\tilde\rho,\varphi,\tilde S_\varphi)\tilde S_{\tilde\rho}
+\tilde C(\eta_0,\tilde\rho,\varphi,\tilde S_\varphi)=0,
\end{equation}
where 
\begin{align}																			\label{eq:4.22}       
&\tilde A(\tilde\rho,\varphi)=\Big(\frac{A}{\rho}-\frac{B}{\rho^2}\rho_0'\Big)^2-1-\frac{{\rho_0'}^2}{\rho^2},
\\
\nonumber
&\tilde B(\eta_0,\tilde\rho,\varphi,\tilde S_\varphi)=\Big(\frac{A}{\rho}-\frac{B}{\rho^2}\rho_0'\Big)
\Big(-\eta_0+\frac{B}{\rho^2}\tilde S_\varphi\Big)+\frac{\rho_0'\tilde S_\varphi}{\rho^2},
\\
\nonumber
&\tilde C(\eta_0,\tilde\rho,\varphi,\tilde S_\varphi)=
\Big(-\eta_0+\frac{B}{\rho^2}\tilde S_\varphi\Big)^2-\frac{\tilde S_\varphi^2}{\rho^2},
\\
\nonumber
&\rho=\tilde\rho+\rho_0'(\varphi).
\end{align}
Since $\tilde\rho =0$  is a characteristic  curve  we have
\begin{equation}																	\label{eq:4.23}             
\tilde A(\tilde \rho,\varphi)=\tilde A_0(\tilde\rho,\varphi)\tilde\rho,
\end{equation}
where  $\tilde A_0(0,\varphi)\neq 0.$

We have
\begin{equation}																	\label{eq:4.24}             
\tilde S_{\tilde \rho}(\tilde\rho,\varphi)=\frac{-\tilde B\pm\sqrt{\tilde B^2-\tilde A\tilde C}}{\tilde A_0(\tilde\rho,\varphi)}.
\end{equation}
The root  $\tilde S_{\tilde\rho}$  that  tends to $\infty$  when 
$\tilde\rho\rw 0$  has the form
\begin{equation}																	\label{eq:4.25}               
\tilde S_{\tilde\rho}(\tilde\rho,\varphi)=\frac{-2\tilde B(\eta_0,0,\varphi,\tilde S_\varphi)}{\tilde A_0(0,\varphi)\tilde\rho}+O(1).
\end{equation}
It follows  from  (\ref{eq:4.20})   that 
$\frac{2B(\eta_0,0,\varphi,\tilde S_\varphi)}{\tilde A_0(0,\varphi)}=\tilde B_1(\varphi)\eta_0+\tilde B_2(\varphi)\tilde S_\varphi$.

As in  (\ref{eq:4.3})  we  approximate  
$\tilde S(\rho,\varphi)$  by the solution  of linear  first order  partial  differential equation
\begin{equation}																	\label{eq:4.26}             
\tilde\rho\tilde S_{1\tilde\rho}(\tilde\rho,\varphi)+\tilde B_1(\varphi)\eta_0+\tilde B_2(\varphi)\tilde S_{1\varphi}(\tilde\rho,\varphi)=0.
\end{equation}
We shall solve (\ref{eq:4.26})
explicitly.  We look  for a particular solution  $\tilde S_2(\tilde \rho,\varphi)$  of nonhomogeneous  equation (\ref{eq:4.26}) 
in the form 
$$
\tilde S_2(\rho,\varphi)=\eta_0b_0\ln\tilde\rho +\tilde S_3(\varphi),
$$
where  $b_0$  will be determined  below,  $\tilde S_3(\varphi)$  is periodic  in  $\varphi$.       For  $\tilde S_3(\varphi)$  we get an equation  
\begin{equation}																	\label{eq:4.27}           
\tilde B_2(\varphi)\tilde S_{3\varphi}(\varphi)+\tilde B_1(\varphi)\eta_0+b_0\eta_0=0.
\end{equation}
A necessary  and sufficient  condition  of the existence  of  a periodic  solution  of  (\ref{eq:4.27})  is
\begin{equation}																	\label{eq:4.28}           
\int\limits_0^{2\pi}\Bigg(\frac{\tilde B_1(\varphi)}{\tilde B_2(\varphi)}+\frac{b_0}{\tilde B_2(\varphi)}\Bigg)d\varphi=0.
\end{equation}
Thus  (\ref{eq:4.28}) 
   determines  $b_0$.  Note that  $\tilde B_2(\varphi)\neq  0$  and
\begin{equation} 																	\label{eq:4.29}          
b_0=\frac{-\int\limits_0^{2\pi} \frac{\tilde B_1(\varphi)}{\tilde B_2(\varphi)}d\varphi}{\int\limits_0^{2\pi}\tilde B_2^{-1}(\varphi)d\varphi}.
\end{equation}
Consider the  homogeneous  equation  (\ref{eq:4.26}):    
\begin{equation}																		\label{eq:4.30}         
\tilde \rho\tilde S_{4\rho}(\tilde\rho,\varphi)+\tilde B_2(\varphi)\tilde S_{4\varphi}(\tilde\rho,\varphi)=0.
\end{equation}
As  in  (\ref{eq:4.3})   the general  solution  of  (\ref{eq:4.25})   
has the form
$$
\tilde S_4(\tilde\rho,\varphi)=g\Big(\ln\tilde\rho -\int\limits_0^\varphi\frac{1}{\tilde B_2(\varphi')}d\varphi'\Big),
$$
where
$g(t)$  is arbitrary.

We choose  $g(t)=b_1t$,  where  $b_1$  is such  that
\begin{equation}																		\label{eq:4.31}    
b_1\int\limits_0^{2\pi}\frac{1}{\tilde B_2(\varphi)}d\varphi =2\pi m,\ \ \ m\in \Z.
\end{equation}
Therefore  finally the solution  of  (\ref{eq:4.26})    
has the form
$$
\tilde S_1(\tilde \rho,\varphi)=(\eta_0 b_0+b_1)\ln\tilde \rho + \tilde S_5(\varphi),
$$
where  $b_0,b_1$  are determined  in  (\ref{eq:4.29})  and  (\ref{eq:4.31})  and  $\tilde S_5(\varphi)$ has  the form
\begin{equation}																		\label{eq:4.32}        
\tilde S_5(\varphi)=\tilde S_3(\varphi)-b_1\int\limits_0^\varphi\frac{1}{\tilde B_2(\varphi')}d\varphi'.
\end{equation}
Note that  $\tilde S_5(\varphi+2\pi)=\tilde S_5(\varphi)+2\pi m,\ m\in  \Z$.

Now we shall  define the wave packet  $\tilde C_0$  as the solution of $\Box_g u=0$ 
 having  the following initial  conditions in  $(\tilde\rho,\varphi)$
coordinates
\begin{align}																		\label{eq:4.33}           
&\tilde C_0\big|_{x_0=0}=\theta(\tilde\rho)\frac{\tilde f(\tilde\rho)}{\sqrt{\rho_0(\varphi)+\tilde\rho}}
e^{i(\eta_0b_0+b_1)\ln\tilde\rho +i\tilde S_5(\varphi)},
\\
\nonumber
&\frac{\partial\tilde C_0}{\partial x_0}\big|_{x_0=0}=i\tilde\beta\theta(\tilde\rho)
\frac{\tilde f(\tilde\rho)}{\sqrt{\rho_0(\varphi)+\tilde\rho}}
e^{i(\eta_0b_0+b_1)\ln\tilde\rho +i\tilde S_5(\varphi)},
\end{align}
where  
\begin{align} 																	\label{eq:4.34}               
&\tilde f(\tilde \rho)=
\tilde\rho^\e e^{-a\tilde \rho},
\\
\nonumber
&\tilde\beta =-\Big(\frac{A(\rho,\varphi)}{\rho} - \frac{B(\rho,\varphi)}{\rho^2}\rho_0'(\varphi)\Big)\frac{(\eta_0b_0+b_1)}{\tilde \rho}
-\frac{B(\rho,\varphi)}{\rho^2}\, \frac{\partial S_5(\varphi)}{\partial \varphi} -\frac{\eta_0 b_0+b_1}{\tilde \rho}.
\end{align}



Let  $\tilde f_k^+(x_0,x),  \tilde f_0^-(x_0,x)$  be  the solutions  of  $\Box_g u=0$  with  the initial conditions  (cf.  (2.1),  (\ref{eq:2.2}))
\begin{equation}																	\label{eq:4.35}      
\tilde f_k^+(x_0,x)\Big |_{x_0=0}=
\tilde\gamma_k e^{i\eta_{\tilde \rho}\tilde \rho +im'\varphi},
\ \ 
\frac{\partial f_k^+}{\partial x_0}\Big|_{x_0=0}=i\tilde \lambda _0^-(k)\tilde\gamma_k e^{i\eta_{\tilde \rho}\tilde\rho +im'\varphi},
\end{equation}
where  (cf.  (\ref{eq:2.3})   and   (\ref{eq:4.13}))
\begin{equation}																		\label{eq:4.36}      
\tilde\lambda_0^-(k)=-\Big(\frac{A}{\rho}-\frac{B}{\rho^2}\rho_0'(\varphi)\Big)\eta_{\tilde \rho}
 -\frac{B}{\rho^2}m'-\sqrt{\eta_{\tilde \rho}^2+\frac{m'^2}{d^2}},
\end{equation}
\begin{equation}																		\label{eq:4.37}          
\tilde\gamma_k=\frac{1}{\sqrt{\rho_0(\varphi)+\tilde\rho}\,\,\big(\eta_{\tilde\rho}^2+\frac{m'^2}{d^2}\big)^{\frac{1}{4}}\sqrt{2(2\pi)^2}},\ \ 
\rho=\rho_0(\varphi)+\tilde\rho,\ \ d \ \ \mbox{is arbitrary}.
\end{equation}
Analogously   (cf.  (\ref{eq:2.4}))
\begin{equation}																	\label{eq:4.38}                  
\tilde f_k^-(x_0,x)\Big |_{x_0=0}=
\tilde\gamma_k e^{i\eta_{\tilde \rho}\tilde \rho +im'\varphi},
\end{equation}
and
$$ 
\frac{\partial f_k^-}{\partial x_0}\Big|_{x_0=0}=i\tilde \lambda _0^+(k)\tilde\gamma_k e^{i\eta_{\tilde \rho}\tilde\rho +m'\varphi},
$$
where $\lambda_0^+(k)$  is similar  to $\lambda_0^-(k)$  with a positive  square root. 
Note that  $\tilde f_k^\pm$  satisfy   ``orthogonality  conditions"  of the forms  (\ref{eq:2.7}),  (\ref{eq:2.8}).



Expanding  $\tilde C_0$ with  respect  to the  basis  $\tilde f_k^{++}, \tilde f_k^{+-},  \tilde f_{-k}^{-+}, \tilde f_{-k}^{--}$
we get, as  in  (\ref{eq:2.17}):
\begin{equation}																		\label{eq:4.39}      
\tilde C_0=\int(
\tilde C^{++}(k)\tilde f_k^{++}+
\tilde C^{+-}(k)\tilde f_k^{+-}-
\tilde C^{-+}(k)\tilde f_{-k}^{-+}-
\tilde C^{--}(k)\tilde f_{-k}^{--})dk,
\end{equation}
where $k=(\eta_\rho,m')$  and  $dk$   means   integration in  $\eta_\rho$  and  the summation  in $m'$  (cf.   (\ref{eq:2.12})   ). 
Note that 
\begin{equation}																		\label{eq:4.40}         
\tilde C^{+-}(\eta_\rho,m')=\tilde C_1^{+-}(k)+\tilde C_2^{+-}(k)
\end{equation}
have  the  same  form  as  (\ref{eq:3.12}),  (\ref{eq:3.13})  with  $\xi_0|A|$  replaced  by  $  (\eta_0 b_0+b_1),
e^{im\varphi} $  replaced  by  $e^{iS_5(\varphi)}$  and  $\rho-|A|$ replaced   by  $\tilde \rho$.

Let, as in  (\ref{eq:4.12}),
\begin{equation}																		\label{eq:4.41}            
\tilde\gamma_{m'}=\frac{1}{2\pi}\int\limits_0^{2\pi} 
e^{-im'\varphi + iS_5(\varphi)}d\varphi.
\end{equation}
If we  replace $\tilde C_0$  by  $\tilde C_n=\frac{\tilde C_0}{<\tilde C_0,\tilde C_0>}$
then  analogously  to  (\ref{eq:4.15})  we  have
\begin{multline}																			\label{eq:4.42}               
\lim_{a\rw\infty}\langle \Psi | N(\tilde C_n) | \Psi\rangle =
\frac{2^{2\e}}{2\pi\Gamma(\e)}
e^{-2\pi (\eta_0b_0+b_1)}|\Gamma_1(\eta_0b_0+b_1)|^2
\frac{(\eta_0b_0+b_1)^2+\e^2}{\eta_0b_0+b_1}
\\
\cdot 
\int\limits_{-\infty}^0
\frac
{|\eta_0|}
{(\eta_\rho^2+1)^{\e+1}}
e^{2(\eta_0b_0+b_1)\sin^{-1}{\frac{1}{\sqrt{\eta_\rho^2+1}}}}d\eta_\rho.
\end{multline}
 Thus we proved  the following theorem:
 \begin{theorem}																		\label{theo:4.2}
 Let  $b_0,b_1$  be the same as in  (\ref{eq:4.29}), 
  (\ref{eq:4.30}). 
   Then  $\langle \Psi | N(\tilde C_n) |\Psi\rangle$  has 
 the form  
(\ref{eq:4.42})  
 when $a\rw\infty$.
\end{theorem}

\section{Conclusion}

In this paper  we study  the Hawking  radiation  for the rotating  acoustic black holes  with 
variable  radial  and  angular  velocities.

In the general  case  of rotating  black holes,  i.e.  when  $A(\rho,\varphi)<0,|B(\rho,\varphi)|>0)$,  
there are  always  black  holes  (see  [6],  [7]).  Some of them  have  a direct  physical  meaning   as it was  shown  in \S 1.

There are two main steps in the  derivation  of the Hawking  radiation:  finding  an appropriate  vacuum  
state  and finding  appropriate vacuum wave
packets,  more  precisely,
sequence of wave  packets,  for the computation of the Hawking radiation. 

The choice  of vacuum state  is not the same as in the Minkowsky space.

We used  a vacuum state  similar to the   one found by W.Unruh  in [23],  [24].  As in [23],  [24]  we split  the set  of the eigenfunctions  of   four 
subsets  instead of two.

Regarding  the wave  packets  we are looking   for  the  wave  packets  having the initial  condition  of the form  
$u=C(\rho)e^{iS(\rho,\varphi)},\ S(\rho,\varphi)$   is a simplified  eikonal   function  and
$$
C(\rho)=\theta(\rho-|A|)\frac{(\rho-|A|)^\e}{\sqrt \rho} e^{-a(\rho-|A|)}.
$$

When the parameter  $a$  increases the wave  packets becomes  closer  and closer   to the black  hole    $\{\rho<|A|\}$.

Assuming that the KG  norm  of the wave packet  is one and taking  taking the limit  when  $a\rw\infty$   we are getting  the Hawking radiation.
Note that   in a rigorous derivation  of the  Hawking radiation in the spherically symmetric case K.Fredenhagen and R.Haag  [11] 
 also used  the limiting procedure when the time $T\rw-\infty$  to obtain    the Hawking radiation.

\end{document}